\documentclass[aps,pra,a4paper,twocolumn,superscriptaddress,floatfix]{revtex4-2}

\usepackage{epsf, latexsym, color, epsfig}
\usepackage{amssymb, amsmath}
\usepackage{amsthm}
\usepackage{times, ulem}
\usepackage{answers}
\usepackage{setspace}
\usepackage{graphicx}
\usepackage{enumitem}
\usepackage{mathrsfs}
\usepackage{cancel}
\usepackage{diagbox}

\usepackage[colorlinks]{hyperref}
\AtBeginDocument{%
	\hypersetup{
		urlcolor     = blue, 
		filecolor   = red,
		linkcolor    = blue, 
		citecolor   = blue, 
	}
} 

\newcommand{\ignore}[1]{}

\newcommand{\ra}{{\rightarrow}}

\newcommand{\be}{\begin{equation}}
\newcommand{\ee}{\end{equation}}
\newcommand{\ba}{\begin{eqnarray}}
\newcommand{\ea}{\end{eqnarray}}

\newcommand{\ket}[1]{| #1 \rangle}

\newcommand{\key}[1]{\begin{center}#1\end{center}}

\begin{document}

\title{Quantum communication networks with optical vortices}

\author{{\c S}erban Suciu}
\affiliation{Horia Hulubei National Institute of Physics and Nuclear Engineering, 077125 Bucharest--M\u agurele, Romania}
\affiliation{Faculty of Physics, University of Bucharest, P.O.Box MG-11, Bucharest-Magurele, Romania}

\author{George Andrei Bulzan}
\affiliation{National Institute for Research and Development in Microtechnologies IMT, Bucharest 077190, Romania}
\affiliation{Faculty of Physics, University of Bucharest, P.O.Box MG-11, Bucharest-Magurele, Romania}

\author{Tudor-Alexandru Isdrail{\u a}}
\affiliation{Horia Hulubei National Institute of Physics and Nuclear Engineering, 077125 Bucharest--M\u agurele, Romania}
\affiliation{Department of Physics, Tamkang University, 151 Yingzhuan Rd., Tamsui Dist., New Taipei City 251301, Taiwan}

\author{Alexandra Maria P{\u a}lici}
\affiliation{Horia Hulubei National Institute of Physics and Nuclear Engineering, 077125 Bucharest--M\u agurele, Romania}
\affiliation{Dept. Photonics Engineering, Technical University of Denmark (DTU), DK-2800 Kgs.~Lyngby, Denmark}

\author{Stefan Ataman}
\affiliation{Extreme Light Infrastructure-Nuclear Physics (ELI-NP), Horia Hulubei National Institute for Physics and Nuclear Engineering, 30 Reactorului Street, 077125 Bucharest--M\u agurele, Romania}

\author{Cristian Kusko}
\affiliation{National Institute for Research and Development in Microtechnologies IMT, Bucharest 077190, Romania}

\author{Radu Ionicioiu}
\affiliation{Horia Hulubei National Institute of Physics and Nuclear Engineering, 077125 Bucharest--M\u agurele, Romania}

\begin{abstract}
Quantum communications bring a paradigm change in internet security by using quantum resources to establish secure keys between parties. Present-day quantum communications networks are mainly point-to-point and use trusted nodes and key management systems to relay the keys. Future quantum networks, including the quantum internet, will have complex topologies in which groups of users are connected and communicate with each-other. Here we investigate several architectures for quantum communication networks. We show that photonic orbital angular momentum (OAM) can be used to route quantum information between different nodes. Starting from a simple, point-to-point network, we will gradually develop more complex architectures: point-to-multipoint, fully-connected and entanglement-distribution networks. As a particularly important result, we show that an $n$-node, fully-connected network can be constructed with a single OAM sorter and $n-1$ OAM values. Our results pave the way to construct complex quantum communication networks with minimal resources.

\key{keywords: quantum networks, OAM, QKD}
\end{abstract}

\maketitle

\section{Introduction}

Quantum computers pose a threat to present-day internet security, due to their ability to efficiently break public-key cryptography. One way to mitigate this {\em quantum apocalypse} is to deploy large-scale quantum communication networks. Current quantum communications are usually point-to-point, using trusted nodes and key management systems to establish secret keys between remote nodes.

Future quantum networks, including the quantum internet, will require to both handle complex network topologies \cite{Frohlich2013} and secure such networks \cite{Yin2023}. These networks will need to connect users situated in different locations and/or domains. Consequently, in such networks it will be important to route a quantum state $\ket{\psi}_q$ between different locations.

Most of the information we exchange everyday is encoded in photons and carried by optical fibres. The data capacity of a single optical fiber depends on the spectral bandwidth over which low-loss signal transmission can be achieved, on the one hand, and on our ability to use this bandwidth through suitable coding and decoding schemes, on the other. 

Due to the constant increase of worldwide data traffic, nonlinear effects \cite{richardson2010} impose limits on the capacity of optical fibers. To address this capacity-crunch, space division multiplexing (SDM) using multi-core \cite{sakaguchi2012, sakaguchi2019} and multimode \cite{ryf2012} fibers have beed developed. In the quest for larger data capacity, another solution is to use an extra degree of freedom, different from wavelength \cite{huang2013, Bozinovic2013}.

A good candidate for the extra degree of freedom is the orbital angular momentum (OAM) of the photon \cite{Terriza2001, andrews2011, Zhou:19}. The phase front of an OAM beam is helical, with quantized angular momentum $l\hbar$, $l\in\mathbb{Z}$. Photons carrying OAM have been used for different applications, such as object identification \cite{Uribe-Patarroyo2013}, enhanced phase sensitivity \cite{DAmbrosio2013}, imaging \cite{Tamburini2006, palici2022} and metrology \cite{Lavery2013, Cvijetic2015}. Classical and quantum communication with OAM states have both been demonstrated in fiber \cite{Bozinovic2013, ingerslev2018, sit2018}, including experimental mode-division multiplexing \cite{Zhang2020}.

Long-distance, high-dimensional QKD using OAM in both optical fibers \cite{Bozinovic2013} and free-space \cite{Sit2017} have recently enjoyed a renewed interest. This is due to several benefits brought by high-dimensional systems: reduced overall complexity of a quantum circuit via $d$-level gates \cite{nikolaeva2022efficient, RalphToffoliQDGates}, increased raw-key rates \cite{SheridanKeyRateQudits, Cerf2002}, robustness to noise \cite{EckerNoise2019, ZhaoNoise, Pasquinucci2000} and hacking attacks \cite{BouchardHacking}. Hybrid states of OAM and polarization have also been used in QKD protocols, in both fiber and free-space \cite{cozzolino2019, DAmbrosio2012, Vallone2014}. Furthermore, recent advances have lowered the resource requirements for point-to-multipoint architectures \cite{Frohlich2013} and have also enhanced quantum digital signatures protocols \cite{Yin2023}. OAM multiplexing can offer an alternative to the development of wireless communications \cite{oam10}, because unlike wavelength-division multiplexing (WDM), it can generate orthogonal channels \cite{oam10} in a line-on-site channel environment.

Due to this increased interest in both classical and quantum applications of OAM, dedicated optical fibers \cite{gregg2015} and multiplexing and demultiplexing techniques \cite{leach2002, OSullivan2012, sorter, Fu2018} have been maturing recently. Thus new methods to route information encoded in OAM are needed. In contrast to wavelength, it is relatively easy to change OAM using passive optical elements like spiral phase-plates (SPPs) \cite{SPPs}. This makes OAM an attractive degree-of-freedom for network routing.

In this paper we discuss several architectures for quantum communication networks which use OAM for routing quantum states $\ket{\psi}_q$ around the network. The paper is structured as follows: in Section \ref{sec:prelim} we describe the quantum sorter \cite{sorter}, which is the main element in OAM multiplexing/demultiplexing (MUX/DEMUX). In Section \ref{sec:pam} we show OAM implementations of several topologies for quantum communications networks: point-to-point, point-to-multipoint, fully connected, and fully-connected entanglement-distribution networks with a central network provider. Finally, we conclude the article in Section \ref{sec:final}.

\section{Quantum sorter}
\label{sec:prelim}

A central element of all the networks discussed here is the $d$-dimensional quantum sorter $U_d$, and its inverse $U_d^{\dagger}$ \cite{sorter}. A short description is provided in Appendix \ref{OAM_sorter}; $U_d$ ($U_d^{\dagger}$) is a unitary operator which acts as a demultiplexer (multiplexer). In quantum information parlance, the sorter $U_d$ is a controlled-$X_d$ gate between the observable to be sorted and the path degree of freedom.

The sorter is universal, i.e., it can (de)multiplex any internal degree of freedom: wavelength, spin, radial angular momentum, OAM etc, and has a theoretical efficiency of 100\% \cite{sorter}. A definite advantage of the sorter is that it can be implemented with linear optics. In Appendix \ref{OAM_sorter} we provide a physical intuition behind the sorter. Experimentally, sorting photons according to their radial number has been realized in Refs.~\cite{Zhou2017,Gu2018}. The same sorting mechanism has been recently applied to a novel method of mass spectrometry \cite{Ionicioiu2023}. Standard telecom networks use wavelength as the extra DoF for multiplexing/demultiplexing. In this article we focus on the OAM degree-of-freedom as a tool for MUX/DEMUX. The action of the sorter $U_d$ and its inverse $U_d^{\dagger}$ is:
\begin{eqnarray}
\label{sorter}
U_d\ (DEMUX): \ \ket{l}_\text{oam}\ket{k}_\text{path} &\xrightarrow{U_d}& \ket{l}_\text{oam}\ket{k\oplus l}_\text{path} \nonumber\\
U_d^{\dagger}\ (MUX): \ \ket{l}_\text{oam}\ket{k}_\text{path} &\xrightarrow{U_d^\dagger}& \ket{l}_\text{oam}\ket{k\ominus l}_\text{path}
\end{eqnarray}
with $\oplus/\ominus$ addition/subtraction $\mathrm{mod}\,d$. Here both OAM and path DoFs are qudits, i.e., $d$-dimensional quantum systems. Thus, if photons with different OAM $l$ are incident on port 0 of the $U_d$ gate (DEMUX), they will exit on output $l$, i.e., they will be sorted on different outputs according to their OAM value. The $U_d^\dagger$ gate (MUX) works in reverse.

The addition and subtraction $\mathrm{mod}\,d$ result in a cyclic property which can be better understood for $l=\pm d$. In this case $\ket{l}_{oam}$ will be sorted on path $\ket{0}_\text{path}$, like $l=0$. This cyclic property will play a crucial role in the design of various network architectures for routing quantum states using OAM.

The cyclic property is also used in the construction of the $X_d$ gate \cite{isdrailua2019cyclic}. The $X_d$ gate is a basic building block for qudit tomography \cite{AsadianTomo, Palici_2020} and for general qudit protocols. Another application of the quantum sorter is in the generation of high-dimensional entangled states between an observable and the path DoF. Hybrid quantum gates are a hot topic under active development \cite{Daboul_Hybrid, KHANHybrid, FicklerHybrid}. Accessing a larger alphabet allows us to encode more information, resulting in higher channel capacity and better robustness to noise.

\section{OAM-assisted quantum communication networks}
\label{sec:pam}

In this section we start with a simple architecture and then gradually build more complex networks. All networks discussed here can be used for QKD, either in prepare-and-measure (BB84) or in entanglement-based protocols (E91, BBM92). The only difference is in the equipment available to users. The networks can also be used to route quantum information as part of a larger protocol. What we denote as "senders" and "receivers" can represent anything from sources and detectors to other networks or protocols. The scale can also vary form waveguides in computer chips to optical fibers between cities or ground-to-satellite links.

For simplicity, in the following we use only positive OAM values. One can substitute the OAM $\ket{l_\text{max}-n}_\text{oam}\mapsto \ket{-n-1}_\text{oam}$, where $l_\text{max}$ is the largest OAM used in the protocol, and $n\in\{0,\;1,\;...,\;\lfloor\frac{l_\text{max}}{2}\rfloor\}$ thus halving the maximum OAM values required.

\subsection{Point-to-point architecture}

Point-to-point networks are a simple case in which pairs of users are connected by their own quantum channel. In practice this results in a messy and convoluted network of cables. To reduce the number of cables needed, especially for long-distance communication, individual signals are in practice multiplexed into the same channel. For example, different labs from two cities can share the same channel for intercity transmission.

\begin{figure}
	\includegraphics[width=\columnwidth]{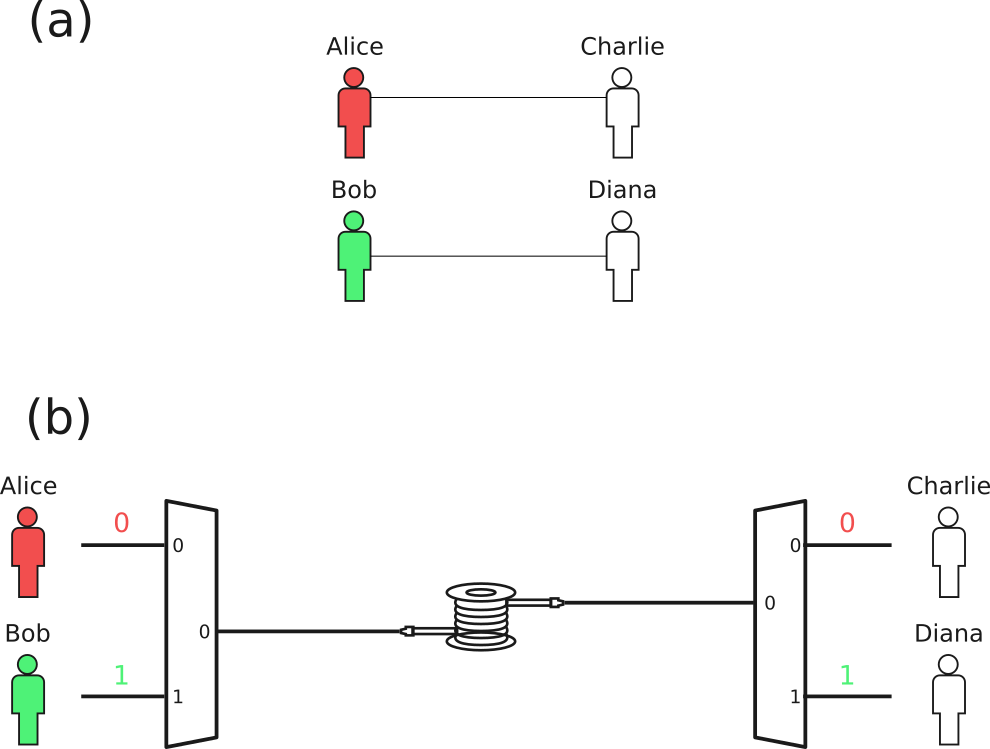}
	\caption{A. Logical network for two pairs connected in a point-to-point topology. B. Two point-to-point pairs using a common channel. Signals from each pair are multiplexed into the common (long-range) channel and demultiplexed at the destination. The information can be recovered and separated because each sender-receiver pair has allocated a unique OAM value.}
	\label{fig:Simple}
\end{figure}

In Figure \ref{fig:Simple} two pairs Alice-Charlie and Bob-Diana share a single long-range quantum channel (instead of dedicated channels for each pair). Each pair has assigned a unique OAM value. The pairs are indexed by consecutive numbers which represents their assigned OAM, input port at the multiplexer and output port at the demultiplexer.

For example, if Bob wants to send a quantum state $\ket{\psi}_q$ to Diana, he uses input port $\ket{1}_\text{path}$ with OAM $\ket{1}_\text{oam}$. This state is input into the multiplexer which redirects it to port $\ket{0}_\text{path}$ of the long-range channel:
\begin{equation}
\ket{\psi}_q \ket{1}_\text{oam} \ket{1}_\text{path} \xrightarrow{MUX} \ket{\psi}_q \ket{1}_\text{oam} \ket{0}_\text{path}\nonumber
\end{equation}

Diana recovers $\ket{\psi}_q$ on output port $\ket{1}_\text{path}$ of the demultiplexer at her end:
\begin{equation}
\ket{\psi}_q \ket{1}_\text{oam} \ket{0}_\text{path} \xrightarrow{DEMUX} \ket{\psi}_q \ket{1}_\text{oam} \ket{1}_\text{path}\nonumber
\end{equation}
The quantum state $\ket{\psi}_q$ can be any internal degree of freedom (different from OAM). Usually we use polarization to encode the quantum state $\ket{\psi}_q= \ket{\psi}_\text{pol}= \alpha\ket{H}+\beta\ket{V}$. In Appendix \ref{BB84} we discuss an example of an OAM-assisted BB84 protocol in polarization.

Since the multiplexer and demultiplexer are modelled by a unitary operation, the protocol also works in reverse. We can reverse the direction in Figure \ref{fig:Simple} such that Charlie and Diana are now the senders and everything works similarly. This is true for all communication protocols discussed here.

\subsection{Point-to-multipoint architecture}

A point-to-multipoint architecture is a natural extension from the point-to-point one. Instead of linking pairs of users, a point-to-multipoint network links a group of users with one or more other groups. However, members of the same group cannot communicate with each-other. The logical network is a bipartite graph.

The simple setup with one multiplexer and one demultiplexer works in this case, but only if the numbers of senders and receivers are coprime (see Appendix \ref{coprime} for a proof). Expanding on the previous example, different labs from two cities can now share not only a transmission line, but also choose to which lab from the other city to send data.

\begin{figure}
	\includegraphics[width=\columnwidth]{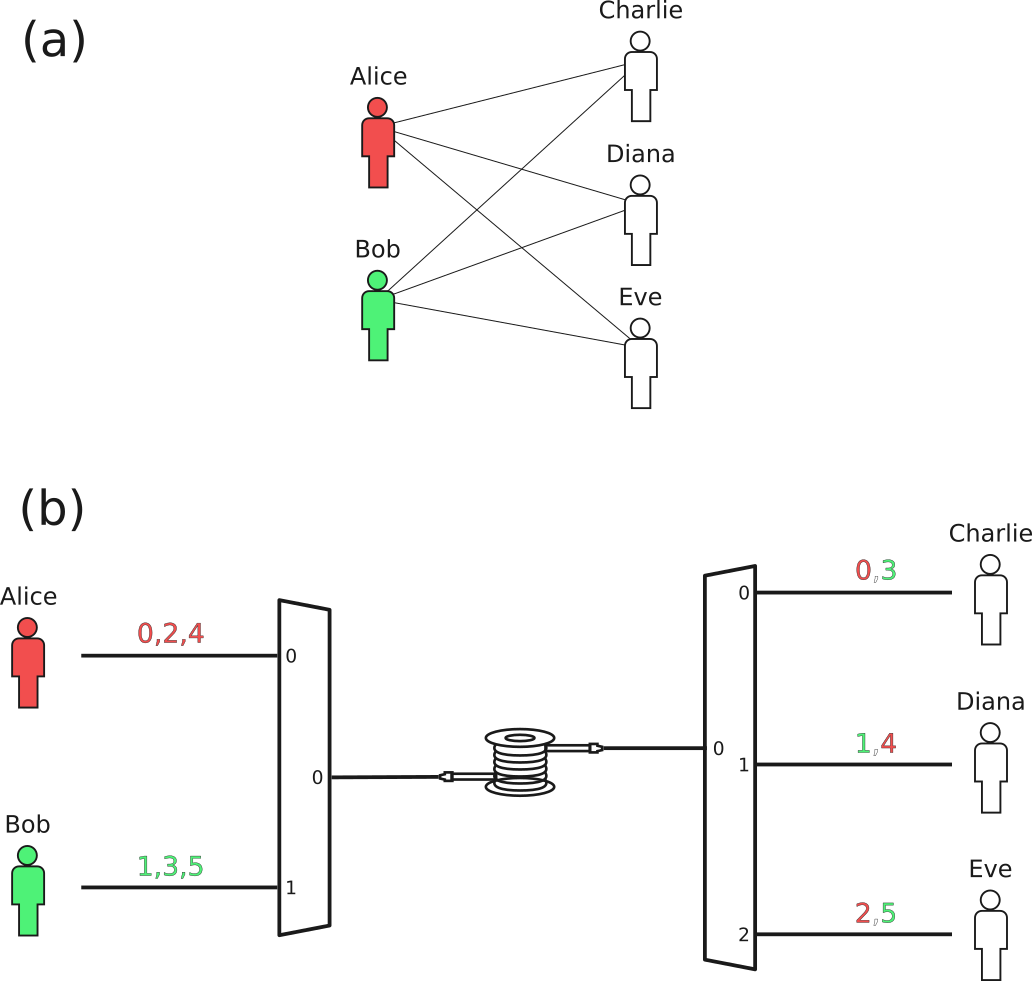}
	\caption{A. Logical network for two groups connected in a point-to-multipoint topology. B. Two groups share a single, long-range channel to communicate with members of the other group. To ensure that each pair has assigned a unique OAM value, the dimensions of multiplexer and demultiplexer must be coprime; here $d_s= 2$ and $d_r= 3$.}
	\label{fig:Coprime}
\end{figure}

Figure \ref{fig:Coprime} shows an example for two senders and three receivers. Alice (Bob) can communicate with any receiver (Charlie, Diana, Eve) using an even (odd) OAM value.

For the general case, suppose we have a set of $d_s$ senders and $d_r$ receivers, with $d_s$, $d_r$ relatively coprime. In this case any sender-receiver pair has associated a unique OAM, thus the receiver can distinguish between different senders. This value is determined by solving a system of congruence relations:
\begin{eqnarray}
	l_{sr}\equiv &s \;(\mathrm{mod}\; d_s)& \nonumber \\
	l_{sr}\equiv &r \;(\mathrm{mod}\; d_r)& \nonumber
\end{eqnarray}
where $d_s$ and $d_r$ are, respectively, the dimension of the multiplexer (sender) and demultiplexer (receiver). In order for a sender $s\in \{0, \ldots, d_s-1 \}$ to communicate with a receiver $r\in \{0, \ldots, d_r-1 \}$, they use the OAM value $l_{sr}$ given by
\be
l_{sr}= p d_s+ s= q d_r+ r
\ee
see Appendix \ref{coprime}. The total number of OAM values is $d_s d_r$.

\begin{figure}[!htb]
	\includegraphics[width=\columnwidth]{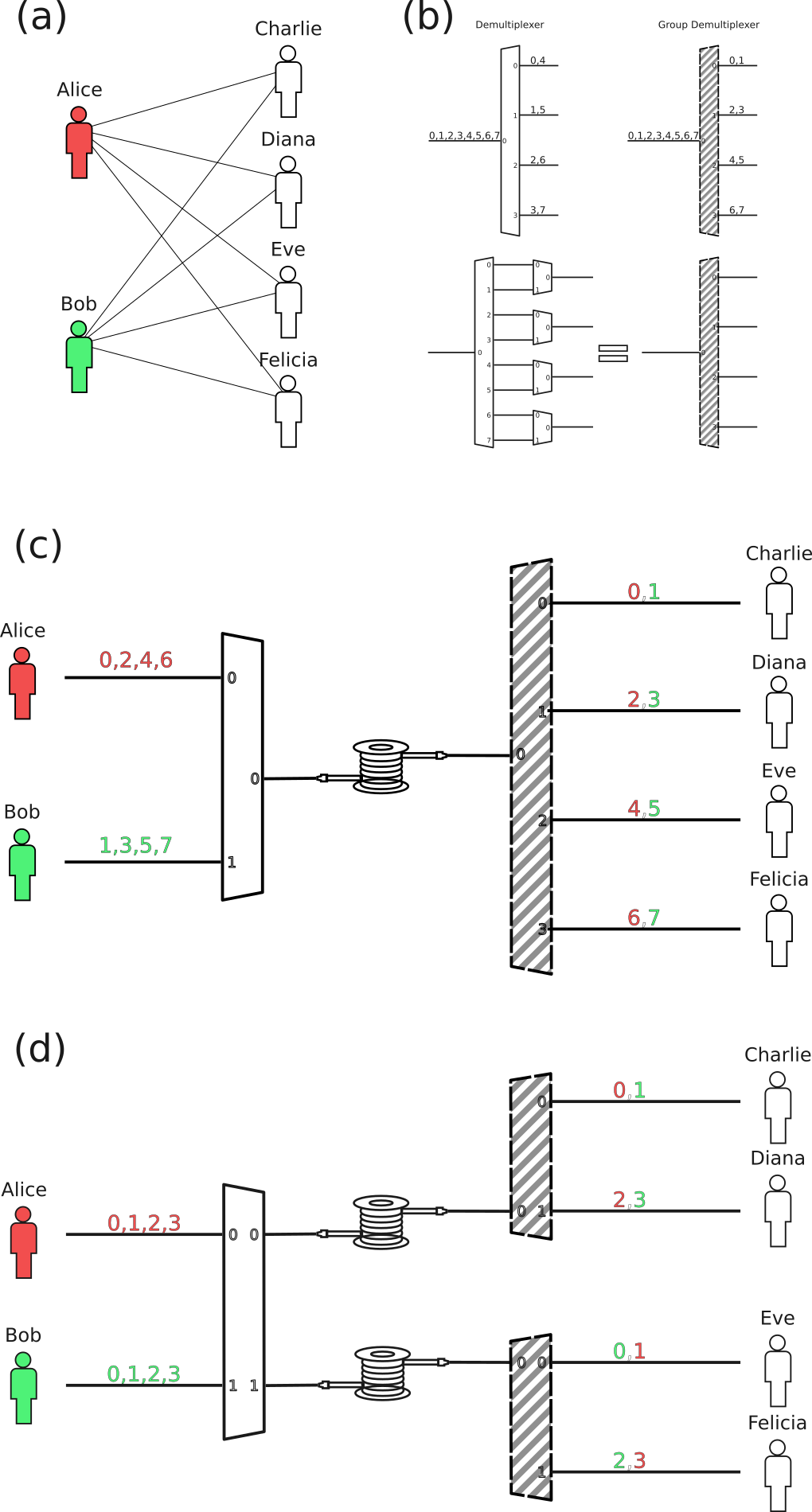}
	\caption{A. Logical network for two groups connected in a point-to-multipoint topology. B. Schematic for a group demultiplexer. C. General point-to-multipoint protocol for an arbitrary number of senders and receivers. On the receiver side, the demultiplexer has been replaced by a group demultiplexer (dashed outline). With this change we eliminate the coprimality condition and the OAM assignment is simplified. D. General point-to-multipoint for multiple groups. Since the multiplexer is a unitary transformation, it has the same number of input and output ports. By using the other available outputs, one group can communicate to several other groups (situated at different locations).}
	\label{fig:General}
\end{figure}

Although in practice we can always choose the dimensions of the multiplexer and demultiplexer to be coprime (e.g., by embedding them into a larger set), this can be an issue for more complex networks. The coprimality constraint can be eliminated by modifying the demultiplexer as in Figure \ref{fig:General} (B). We call this a {\em group demultiplexer} since it splits an input channel into $d_s \cdot d_r$ outputs and then groups them back together into $d_r$ channels.

In Figure \ref{fig:General} (C) we use the group demultiplexer to create a more general network. Now any sender $s$ can transmit to any receiver $r$ by an appropriate OAM value $l_{sr}$:
\begin{equation}
l_{sr}= s+ rd_s
\end{equation}
A step-by-step analysis of this protocol is given in Appendix \ref{m2m}; here we only give the main result:
\begin{equation}
\ket{\psi}_q \ket{l_{sr}}_\text{oam} \ket{s}_\text{path} \xrightarrow{\text{network}} \ket{\psi}_q \ket{l_{sr}}_\text{oam} \ket{r}_\text{path} \nonumber
\end{equation}
This ensures that quantum information, encoded in the state $\ket{\psi}_q$, is routed along the network from sender $s$ to receiver $r$.

In the following schemes a group demultiplexer can be replaced by a simple demultiplexer, provided that: (i) the number of senders and receivers are co-prime, and (ii) the OAM values satisfy the congruence relations discussed above. Also, any network can work in reverse, i.e., receivers become senders, multiplexers become demultiplexers (and vice-versa) and group demultiplexers become group multiplexers.

In point-to-multipoint networks, a group communicates with multiple other groups. A useful use-case scenario is a network connecting multiple cities: labs in one city communicate to labs in multiple cities. However, one network connects just one city with others. This creates a physical star-network topology, the logical network topology remains the same. Each sender forms a logical star-network topology with all receivers, yet as a group the point-to-multipoint logic is unchanged.

In Figure \ref{fig:General} (D) we split de group demultiplexer at the receivers end into two. In large networks it will be useful to put a group multiplexer at the senders and have simple demultiplexers at the receivers. This helps to reduce the costs, since using group multiplexers scales as $d_s\cdot d_r$.

Compared to the previous protocol, we now use other outputs of the multiplexer to communicate with different groups. Everything remains the same, except from an offset of the OAM value, which depends on the receiver group:
\begin{equation}
l_{sgr}= s+ rd_s \ominus g
\end{equation}
where $g$ is the group number (i.e., the output port of the multiplexer).

Consider the example in Figure \ref{fig:General} (D), where Bob intends to communicate to Eve; we have sender $s=1$ transmitting to receiver $r=0$ from group $g=1$, with a multiplexer of size $d_s=2$. Thus their OAM value is $0$. Notice that this OAM value is no longer unique, since Alice uses the same OAM to communicate to Charlie.

This architecture helps to reduce the OAM bandwidth, i.e., the number of OAM values required. Both schemes in Figure \ref{fig:General}, have $d_s=2$ senders and $d_r=4$ receivers. However, in case C we need $d_s d_r= 6$ OAM values, whereas in case D we need only $d_r=4$ values. In both cases any sender can communicate with any receiver.

A variation of this architecture is to use a group multiplexer at the senders side and only demultiplexers at the receivers side. In this case the OAM value is
\begin{equation}
l_{sgr}= r+ sd_r \ominus gd_s
\end{equation}

\subsection{Fully-connected networks}

Finally, we generalize the previous schemes to a fully-connected network, in which any two users can communicate with each other. In this case all nodes are both senders and receivers. In the previous point-to-multipoint protocol, this will work for a reasonable numbers of users, as the size of the group demultiplexer scales as $O(n^2)$. 

\begin{figure}
	\includegraphics[width=\columnwidth]{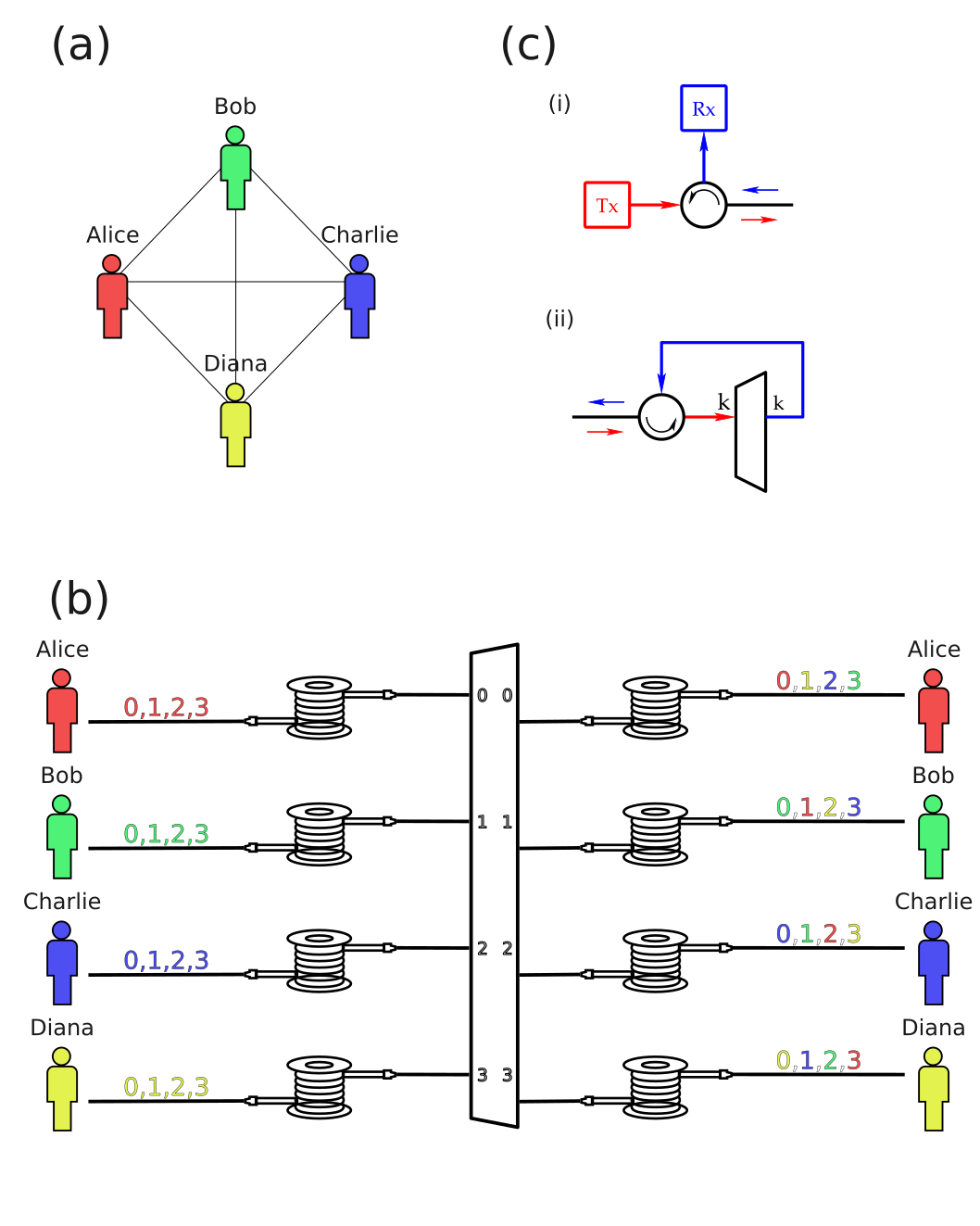}
	\caption{A. Logical network for a fully connected network topology. B. A fully-connected, $n$-node quantum network; each node is both a sender and a receiver. A single $n$-dimensional sorter (MUX/DEMUX) routes information between all the nodes. C. Circulators on the node side (a), and DEMUX side (b), allow each node to be both a transmitter Tx and a receiver Rx. Each node is connected to the (central) DEMUX by a single fiber, and thus to rest of the network.} 
	\label{fig:FullNetSimple}
\end{figure}

Surprisingly however, a single MUX/DEMUX is enough to create a fully-connected network with $n$ nodes, see Figure \ref{fig:FullNetSimple}. Here the senders can view the OAM value as an indexing list: $0$ for themselves, $1$ for the next user (mod $d$), $2$ for the second over (mod $d$) and so on. Now each node is both a sender and a receiver.

\begin{center}
	\begin{table}[h]
		\caption{\label{tab:fully_connected} OAM assignments for the fully-connected network in Figure \ref{fig:FullNetSimple}. Each row is obtained from the one above by a circular right-shift.}
		\begin{tabular}{c|cccc}
			\diagbox{sender}{receiver}&
			A & B & C & D \\
			\colrule\\
			A & $\ket{0}$ & $\ket{1}$ & $\ket{2}$ & $\ket{3}$ \\[.5em]
			B & $\ket{3}$ & $\ket{0}$ & $\ket{1}$ & $\ket{2}$ \\[.5em]
			C & $\ket{2}$ & $\ket{3}$ & $\ket{0}$ & $\ket{1}$ \\[.5em]
			D & $\ket{1}$ & $\ket{2}$ & $\ket{3}$ & $\ket{0}$ \\[.5em]
		\end{tabular}
	\end{table}
\end{center}

Thus a fully-connected network with $n$ users requires only a single $n$-dimensional quantum sorter (acting as a MUX/DEMUX) and $n$ OAM values. In fact, since the OAM value $\ell= 0$ is used to connect a node to itself, we need only $n-1$ OAM values.

At first sight it looks like each node needs two different channels to connect to the network, one for sending and one for receiving. However, using two circulators, one at the user side and the other at the MUX/DEMUX side, a node can use a single channel for both sending and receiving, see Figure \ref{fig:FullNetSimple} (C).

This design is the most general possible, connecting all network nodes. It can be used as a fully-connected network in prepare-and-measure QKD. The strength of this approach can be seen in Table \ref{tab:scaling}, since the fully-connected network has the lowest resource requirements.

\begin{center}
\begin{table}[h]
	\caption{Resource scaling for different network architectures. The fully connected network requires one quantum sorter, whereas all the other architectures require at least two.}
	\label{tab:scaling}
	\begin{tabular}{c|c}
		Network architecture & Resource scaling  \\
		\colrule\\
		Point-to-point & 1$\times U_{d_s}$; 1$\times U_{d_r}$  \\[.5em]
		Point-to-multipoint (general) &  1$\times U_d$; $\left(\frac{d}{2}+1\right) \times U_2$  \\[.5em]
		Point-to-multipoint (groups) & 1$\times U_{d_s}$; $d_s \times U_{d_r}$  \\[.5em]
		Fully Connected & 1$\times U_d$  \\[.5em]
		Entanglement distrib.~(active) & 1$\times U_d$; $\left(\frac{d}{2}+1\right) \times U_2$  \\[.5em]
		Entanglement distrib.~(passive) & 3$\times U_d$; $2d \times$ SPP($i$)
	\end{tabular}
\end{table}
\end{center}

\subsection{Entanglement-distribution networks}

So far we have discussed networks for prepare-and-measure QKD protocols, such as BB84. Another important class of QKD protocols are entanglement-based ones, e.g., E91 or BBM92. Although entanglement-based protocols are more secure than prepare-and-measure ones, they are also more difficult to implement, as they require to distribute entanglement between nodes.

Multi-user entanglement-distribution networks have been experimentally demonstrated for wavelength multiplexing \cite{Entnet, Trunet}. Similar passive-switching networks with a central node can be designed for OAM.

Entanglement-distribution networks can be actively- or passively switched. In an active network, the central node (the source) generates pairs of polarization-entangled photons. An active switch then assigns the correct OAM values $r$ and $s$ to the two photons and then distributes the photons to the corresponding nodes $r$ and $s$.

\begin{figure}
	\includegraphics[width=\columnwidth]{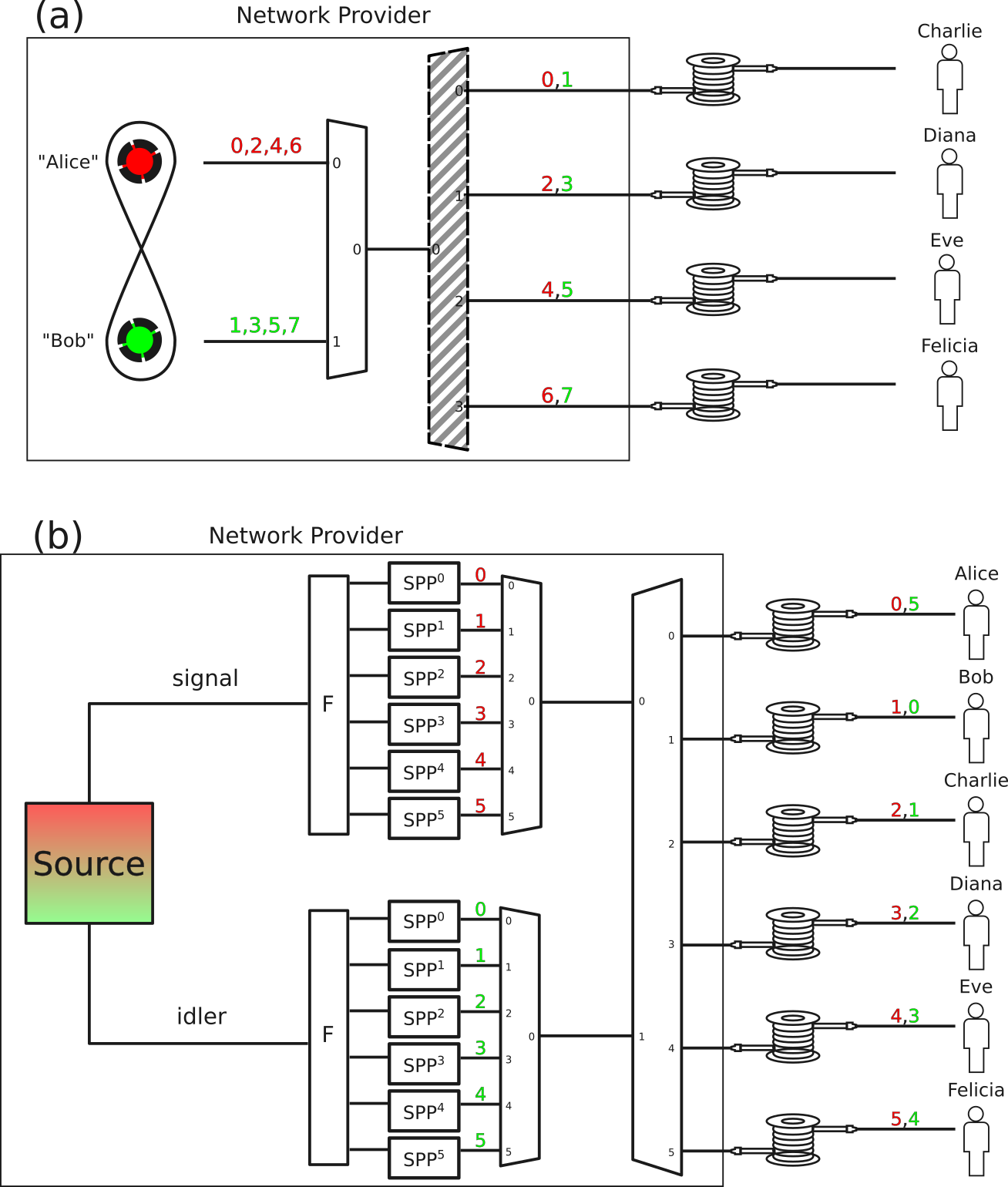}
	\caption{A. Fully-connected network for entanglement distribution using an active central provider. If in the general point to multipoint network one group sends entangled pairs (or multipartite entangled states) to the other, the receiving group becomes a fully-connected network for entanglement-based QKD. B. Fully-connected network for entanglement distribution from a passive central provider. In this scheme the senders are replaced by a source of entangled photons which are then randomly assigned OAM numbers, thus generating random pairs of users which share an entangled state.}
	\label{fig:Entangled}
\end{figure}

For example, the actively-switched entanglement distribution scheme in Ref.~\cite{Entnet} can be translated directly in the OAM domain with a general point-to-multipoint network, as in Figure \ref{fig:Entangled} (A). Here "Alice" and "Bob" are two entangled photons that are distributed based on their assigned OAM value.

Notice that even though the network is not fully connected in the prepare-and-measure regime, it becomes so for entanglement distribution. It is fully connected, i.e., any two users can share an entangled pair. A passive-switching network for entanglement distribution has been experimentally demonstrated in Ref.~\cite{Trunet}. 

A passively-switched network works in a similar way to the choice of measurement basis in BB84, where a beam-splitter chooses randomly the basis. In the case of a passive OAM network, two Fourier gates $F_d$ (which generalize the beamsplitter for $d>2$) put the signal and idler photon into two distinct multi-path interferometers, one for signal and one for idler, see Figure \ref{fig:Entangled} (B). For the signal photon we have:
\begin{equation}
	F_d\ket{0}_\text{path}=\frac{1}{\sqrt{d}}\sum_{i=0}^{d-1}\ket{i}_\text{path} \nonumber
\end{equation}

Each path $i$ of the interferometer has an $i$-th order spiral phase plate (SPP) and changes the OAM $\ket{0}_\text{oam} \ra \ket{i}_\text{oam}$
\begin{equation}
\text{SPP}^i \ket{0}_\text{oam} \ket{i}_\text{path} = \ket{i}_\text{oam} \ket{i}_\text{path} \nonumber
\end{equation}

The paths (channels) are then multiplexed into a common exit path. Subsequently, the two photons are input into path 0 (the signal) and, respectively 1 (the idler) of a final OAM demultiplexer, which distributes the two photons to the final users, Figure \ref{fig:Entangled} (B). The final quantum state of the two photons is (for simplicity we omit the polarization part):
\begin{equation}
\frac{1}{d} \sum_{i=0}^{d-1} \sum_{j=0}^{d-1} \ket{i, j\ominus 1}_\text{oam} \ket{i, j}_\text{path}	
\end{equation}

In this case entanglement distribution between nodes is done randomly, according to the OAM values of signal and idler (via post-selection). Similar to other passively-switched networks \cite{Trunet}, the pairs of nodes (randomly) receiving the entangled pair are identified by coincidences in their detectors.

\subsection{Non-ideal case}

So far we have discussed the ideal, noiseless case. We now briefly analyse the effect of noise. There are two types of losses: (i) in the quantum channel (optical fibres, free-space, underwater etc); (ii) in the quantum sorters. Losses in the quantum channel depend on the specific losses in the fibers (e.g., Raman absorption), atmospheric turbulence, water turbidity etc . Since these are common to all quantum communication protocols using the same type of channel, we will not discuss them here.

Losses due to a non-ideal mass sorter have been discussed in Ref.~\cite{Ionicioiu2023}, including a discussion about decoherence. The OAM sorter and mass sorter have the same quantum network: they are both equivalent to a controlled-$X_d$ gate $C(X_d)$. Thus the conclusion of the previous analysis also holds for the OAM sorter. Specifically, for $d= 3$ the probability of sorting correctly is $>96\%$ even for phase errors as high as $2\pi/15$, representing 20\% of the relevant phase for the system, see Fig.6 of Ref. \cite{Ionicioiu2023}.

\section{Discussion and conclusions}
\label{sec:final}

The development of quantum communication networks and the advent of the future quantum internet is contingent on the ability to route quantum information in networks with complex topologies. In the scenario investigated here, a set of users send quantum states $\ket{\psi}_q$, entangled or not, to another set of receivers. Similar to the classical case, the scarcity of certain resources, like long-distance optical cables, means that signals between different nodes use a common communication channel. This implies that we need to multiplex/demultiplex quantum signals from/to different users (nodes). In order to achieve this, here we use OAM to route the quantum state $\ket{\psi}_q$ between different users.

In this article we have discussed several network architectures for pairwise communication between multiple parties. Starting from a simple, one-to-one network, we then developed one-to-many and fully-connected networks for distributing quantum states. We have shown that a fully-connected network with $n$ nodes can be achieved with minimal resources: a single quantum sorter acting as a MUX/DEMUX, connected to all $n$ nodes. Moreover, this fully-connected network requires only $n-1$ OAM values. Finally, we have developed a novel entanglement-distribution protocol which has several advantages compared to the current wavelength-based networks.

The central element of all the networks discussed here is the quantum sorter which acts as a MUX/DEMUX. The quantum sorter has a cyclic property which was used extensively in building a multitude of network architectures. These networks can be used to distribute quantum states between nodes, both in prepare-and-measure protocols (BB84) and in entanglement-based ones (E91, BBM92).

The protocols described here can be implemented either in optical fibers or in free-space. As future applications, we envisage our protocols to be used in a wide-range of communication tasks, such as terrestrial networks (intra- and inter-city), satellite-to-satellite or satellite-to-ground.

\begin{acknowledgments}
The authors acknowledge support from a grant of the Romanian Ministry of Research and Innovation, PCCDI-UEFISCDI, project number PN-III-P1-1.2-PCCDI-2017-0338/79PCCDI/2018, within PNCDI III. R.I.~acknowledges support from PN 19060101/2019-2022. S.A.~acknowledges the support of contract PN 23 21 01 05 funded by the Romanian Ministry of Research, Innovation and Digitalization and by the Extreme Light Infrastructure Nuclear Physics Phase II, a project co-financed by the Romanian Government and the European Union through the European Regional Development Fund and the Competitiveness Operational Programme (No. 1/07.07.2016, COP, ID 1334). G.A.B.~and C.K.: this research was funded by the Ministry of Research, Innovation and Digitization, CNCS-UEFISCDI via project LACAS grant number PN-III-P2-2.1-PED-2021-1233. T.A.I.~is supported by National Science and Technology Council, Taiwan, under Grant no.~NSTC 112-2112-M-032-008-MY3, 112-2811-M-032-002-MY3 and 111-2923-M-032-002-MY5.
\end{acknowledgments}

\bibliography{ArticolComm}

\appendix
\section{Quantum OAM sorter}
\label{OAM_sorter}

\begin{figure}
	\includegraphics[width= .8\columnwidth]{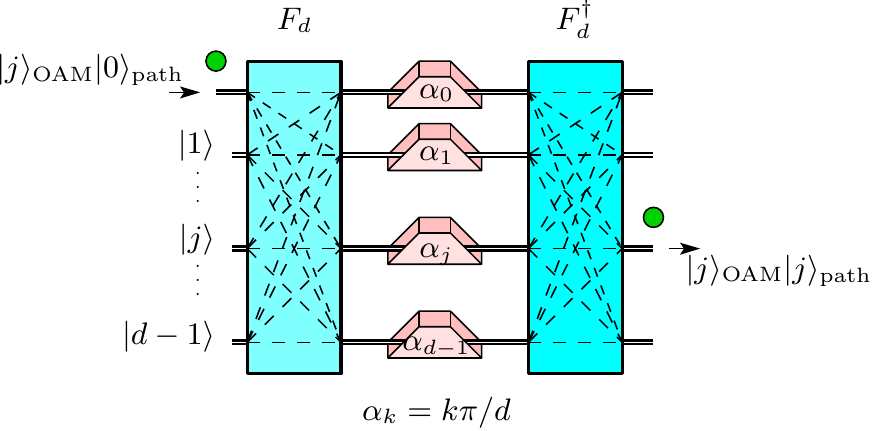}
	\caption{The OAM sorter $U_d$ is a Mach-Zehnder interferometer with $d$ paths and with different phase-shifts $\alpha_k$ on each path. A photon with OAM $\ket{j}$ enters on path $0$ and exits through path $\ket{j}$ with unit probability, i.e., is sorted according to its OAM value.}
	\label{fig:S_oam}
\end{figure}

The OAM sorter $U_d$ used here is an example of the more general universal quantum sorter introduced in \cite{sorter}. The sorter is a Mach-Zehnder interferometer with $d$ paths and with different phase-shifts in each arm. For path $k$, the phase-shifts are given by Dove prisms rotated with angles $\alpha_k= k\pi/d$, $k=0,\ldots , d-1$, see Figure \ref{fig:S_oam}. The $F_d, F_d^\dag$ are discrete Fourier gates acting only on the path degree of freedom; they are equivalent to multi-mode couplers with appropriate phases:
\begin{equation}
	F_d\ket{k}_\text{path}= \frac{1}{\sqrt{d}} \sum_{j=0}^{d-1} \omega^{kj} \ket{j}_\text{path} \nonumber
\end{equation}
with $\omega= e^{2\pi i/d}$ a root of unity of order $d$. We also have $F_d^4=I, F_d^\dag= F_d^3$ and $F_d^2\ket{k}= \ket{-k}= \ket{d-k}$.

Due to constructive interference, a particle with OAM $\ket{j}$ entering the interferometer on input $0$, will exit with unit probability on path $j$: $\ket{j}_\mathrm{oam} \ket{0}_\mathrm{path} \rightarrow \ket{j}_\mathrm{oam} \ket{j}_\mathrm{path}$.

This might seem quite abstract. We now give a physical intuition of how the sorter works. Consider first the simplest case of sorting two states, $d=2$. In this case the sorter is a Mach-Zehnder interferometer with a Dove prism in each arm; the prisms are rotated relative to each other by $\pi/2$ \cite{leach2002}. Thus, an incoming photon with $\ell= 0$ will get a relative phase-shift (between the two arms) $\delta \varphi= 0$ and will exit, with unit probability, through output port 0 (constructive interference). On the other hand, an incoming photon with $\ell= 1$ will get a relative phase-shift $\delta \varphi= \pi/2$ and will exit through output port 1, again with unit probability (destructive interference).

The general case is similar. The sorter is now a Mach-Zehnder interferometer with $d$ paths and its role is to induce state-dependent relative phases between the arms. For example, a photon in the state $\ell= 0$ will have relative phases $(0, 0, \ldots, 0)$ and will exit with unit probability on exit 0. A photon with $\ell= 1$ will have relative phases $(0, \pi/d, \ldots, (d-1)\pi/d)$ and will exit with unit probability on exit 1, and so on. Thus the relative phases experienced by a photon depend on the OAM value $\ell= k$. In turn, the relative phases determine, through constructive/destructive interference, the output $k$ through which the photon will exit.

\section{BB84}
\label{BB84}

The OAM-assisted BB84 protocol in polarization encoding can be implemented with the network shown in Figure \ref{fig:Simple}. In this case there are $d$ senders and receivers.  Sender $i$ sends to receiver $i$ the qubit $\ket{\psi_i}_q:= \alpha_i\ket{H}+\beta_i\ket{V}$, encoded in polarization. Each pair (sender $i$, receiver $i$) has allocated an OAM value $i \in \{ 0, \ldots, d-1 \}$, thus the initial state at sender $i$ is
\begin{equation}
\ket{\psi_i}_q \ket{i}_\text{oam}\ket{i}_\text{path} \nonumber
\end{equation}
Since the polarization qubit is unchanged by the MUX/DEMUX network, we will thus omit it for simplicity. The action of the device is
\begin{equation}
\label{fullaction}
\ket{i}_\text{oam}\ket{i}_\text{path} \xrightarrow{U_d^\dagger}\ket{i}_\text{oam}\ket{0}_\text{path} \xrightarrow{U_d} \ket{i}_\text{oam}\ket{i}_\text{path} \nonumber
\end{equation}

In order to measure the polarization state, each receiver has the standard BB84 setup: a beam-splitter (BS), a half-wave plate (HWP), two polarizing beam-splitters (PBS) and 4 single-photon detectors (SPDs).

The protocol also works for the networks discussed in Section \ref{sec:pam}, if the sender chooses an appropriate OAM value. In Figure \ref{fig:General}, the sender $s$ choses an OAM value equal to $r$ in order to send to receiver $r$ (see Appendix \ref{m2m}):
\begin{equation}
\ket{\psi_{sr}}_q \ket{r}_\text{oam}\ket{s}_\text{path} \nonumber
\end{equation}

In Figure \ref{fig:Coprime}, the sender $s$ communicates with receiver $r$:
\begin{equation}
\ket{\psi_{sr}}_q \ket{sqd_r+rpd_s}_\text{oam}\ket{s}_\text{path} \nonumber
\end{equation}
where $p$ and $q$ are the Bezout coefficients of the identity $pd_s+qd_r=1$ and $d_s$ and $d_r$ are the number of senders and receivers respectively (see Appendix \ref{coprime}).

\section{Co-prime case}
\label{coprime}

For $d_s$ senders and $d_r$ receivers we can prove that any sender $s$ can transmit to any receiver $r$ if they share an OAM state $\ket{l_{sr}}_\text{oam}$ satisfying the conditions
\begin{eqnarray}
l_{sr}= &pd_s+s& \nonumber \\
l_{sr}= &qd_r+r& \nonumber
\end{eqnarray}
with $s\in\{0, \ldots, d_s-1 \}$, $r\in\{0, \ldots, d_r-1 \}$ and $p$, $q$ integers. This is equivalent to the following congruence relations
\begin{eqnarray}
l_{sr}\equiv &s \;(\mathrm{mod}\; d_s)& \nonumber \\
l_{sr}\equiv &r \;(\mathrm{mod}\; d_r)&
\label{congruence} 
\end{eqnarray}

From the Chinese remainder theorem we know that $d_s$ and $d_r$ need to be co-prime and only one $l_{sr} \in \{0,\ldots,d_sd_r-1\}$ satisfies these conditions for fixed $s$ and $r$. This means that we can design a network with only a $d_s$-dimensional MUX $U_{d_s}^\dag$ and a $d_r$-dimensional DEMUX $U_{d_r}$, see Figure \ref{fig:Coprime}.

Given a sender $s$ and a receiver $r$, we can find their assigned OAM state $\ket{l_{sr}}_\text{oam}$ by solving the congruence relations (\ref{congruence}). This gives $l_{sr}= sqd_r+rpd_s$, where $p$ and $q$ are the Bezout coefficients of the identity $pd_s+qd_r=1$, which are calculated using the extended Euclidean algorithm. 

The protocol requires the number of senders and the number of receivers to be coprime. In practice we can always satisfy the co-primality condition by embedding the number of senders and/or receivers into larger sets with co-prime cardinality.

\section{General point-to-multipoint OAM network}
\label{m2m}

Our goal is to establish a general pairwise communication protocol between $d_s$ senders and $d_r$ receivers, such that any sender $s$ can communicate with any receiver $r$. The protocol must be free from the co-primality condition discussed in Appendix \ref{coprime}. In this case there are $d_sd_r$ pairs (sender $s$, receiver $r$), with $s \in\{0, \ldots, d_s-1\}$ and $r\in\{0, \ldots, d_r-1 \}$.

We group the OAM values as in Table \ref{tab:chart}, where each line shows the OAMs available to sender $s$, and each column the OAMs received by receiver $r$.
\begin{table}[htpb]
	\caption{\label{tab:chart}
		OAM correspondence chart for $d_s$ senders and $d_r$ receivers.}
	\begin{tabular}{c|cccc}
		\diagbox{sender}{receiver}&
		$0$ & $1$ & ... & ${d_r-1}$ \\
		\colrule\\
		$0$ & $\ket{0}$ & $\ket{d_s}$ & ... & $\ket{(d_r-1)d_s}$ \\[.5em]
		$1$ & $\ket{1}$ & $\ket{d_s+1}$ & ... & $\ket{(d_r-1)d_s+1}$ \\[.5em]
		... & ... & ... & ... & ... \\[.5em]
		${d_s-1}$ & $\ket{d_s-1}$ & $\ket{2d_s-1}$ & ... & $\ket{d_rd_s-1}$
	\end{tabular}
\end{table}

Notice that the element on row $s$, column $r$ is $\ket{s+rd_s}$, thus we can assign each pair (sender $s$, receiver $r$) a unique OAM state $\ket{l_{sr}}_\text{oam}$, with $\;l_{sr}=s+rd_s$, requiring $d_sd_r$ OAM states in total.

Example: For two senders ($d_s=2$) and three receivers $(d_r=3)$, if sender $0$ ($s=0$) wants to communicate with receiver $1$ ($r=1$), the appropriate OAM state is $\ket{l_{sr}}_\text{oam}=\ket{s+rd_s}_\text{oam}=\ket{0+1\cdot 2}_\text{oam}=\ket{2}_\text{oam}$, see Table~\ref{tab:example}.

\begin{table}[htpb]
	\caption{\label{tab:example}
		Reference chart for $d_s=2$, $d_r=3$ and OAM DoF spanning from $\ket{0}$ to $\ket{5}$.}
		\begin{tabular}{c|ccc}
			\diagbox{sender}{receiver}&
			0 & 
			1 &  
			2 \\
			\colrule\\
			0& $\ket{0}$ & $\ket{2}$ & $\ket{4}$ \\[.5em]
			1 & $\ket{1}$ & $\ket{3}$ & $\ket{5}$ \\[.5em]		
		\end{tabular}
\end{table}

Figure \ref{fig:General} (C) shows the physical implementation of the general point-to-multipoint network for two senders and four receivers. In general, sender $s$ starts with the OAM state $\ket{s+rd_s}_\text{oam}$ on path $\ket{s}_\text{path}$. The state is then multiplexed into a single transmission channel
\begin{equation}
\ket{s+rd_s}_\text{oam}\ket{s}_\text{path} \xrightarrow{U_{d_s}^\dagger} \ket{s+rd_s}_\text{oam}\ket{0}_\text{path} \nonumber
\end{equation}

The demultiplexer at the receiver's end is much larger, splitting the channel into $d_sd_r$ paths, one for each OAM. These are then grouped into $d_r$ groups of $d_s$ channels and multiplexed back together by $d_r$ multiplexers $U_{d_s}^{\dagger}$. For simplicity, we can take the demultiplexer and the $d_r$ multiplexers to be a single device called a {\em group demultiplexer} $G$, represented by a dashed outline in the figures (see Appendix \ref{group}). We mark the paths inside the device as local paths, therefore
\begin{eqnarray}
	\ket{s+rd_s}_\text{oam}\ket{0}_\text{path} &\xrightarrow{U_{d_sd_r}}& \ket{s+rd_s}_\text{oam}\ket{s+rd_s}_\text{local} \nonumber \\
	\ket{s+rd_s}_\text{oam}\ket{s+rd_s}_\text{local} &\xrightarrow{U_{d_s}^\dagger}& \ket{s+rd_s}_\text{oam}\ket{rd_s}_\text{local} \nonumber
\end{eqnarray}

There are gaps of $d_s$ between output ports on the local path. We can map back to the global path by dividing the output port by $d_s$ ($\ket{rd_s}_\text{local}\mapsto \ket{r}_\text{path}$). The full action of the group demultiplexer $G^{d_r}_{d_s}$ (makes $d_r$ groups of dimension $d_s$) is then
\begin{equation}
	\ket{s+rd_s}_\text{oam}\ket{0}_\text{path} \xrightarrow{G^{d_r}_{d_s}}\ket{s+rd_s}_\text{oam} \ket{r}_\text{path} \nonumber
\end{equation}

Finally, the action of the protocol is
\begin{equation}
\ket{s+rd_s}_\text{oam}\ket{s}_\text{path} \xrightarrow{U_{d_s}^\dagger,\; G^{d_r}_{d_s}} \ket{s+rd_s}_\text{oam}\ket{r}_\text{path} \nonumber
\end{equation}

If sender $s$ wants, for example, to transmit a qubit  to receiver $r$, they encode the information in the polarization dof $\alpha_{sr}\ket{H}+\beta_{sr}\ket{V}$ with OAM $\ket{s+rd_s}_\text{oam}$ on path $\ket{s}_\text{path}$. The receiver recovers the information encoded in the qubit via polarisation state detection on path $\ket{r}_\text{path}$. This way we can ensure the general pairwise quantum communication between several parties through a single channel. Moreover, if the two groups of senders and receivers are the same, the network becomes effectively a fully-connected network.

\section{Group demultiplexer}
\label{group}

In Figure \ref{fig:General} (B), we introduce the group demultiplexer $G$. Imputing consecutive OAM numbers on port $\ket{0}_\text{path}$ into a demultiplexer $U_d$, we see that it distributes them on consecutive channels $l\;\text{mod}\;d$ until it resets to output $0$ when $l$ reaches the next multiple of $d$. A group demultiplexer $G^{d_r}_{d_s}$ in the same situation outputs on the same channel $\lfloor\frac{l}{ds}\rfloor$ until it moves to the next one when $l$ reaches the next multiple of $d_s$ as in Table \ref{tab:group}.

\begin{table}[htpb]
	\caption{\label{tab:group} Outputs of a group demultiplexer.}
		\begin{tabular}{c|cccc }
			\diagbox{output}{OAM}&
			$0$ & $1$ & ... & ${d_s-1}$ \\
			\colrule\\
			$0$ & $\ket{0}$ & $\ket{1}$ & ... & $\ket{d_s-1}$ \\[.5em]
			$1$ & $\ket{d_s}$ & $\ket{d_s+1}$ & ... & $\ket{2d_s-1}$ \\[.5em]
			... & ... & ... & ... & ... \\[.5em]
			${d_r-1}$ & $\ket{(d_r-1)d_s}$ & $\ket{(d_r-1)d_s+1}$ & ... & $\ket{d_rd_s-1}$
		\end{tabular}
\end{table}

The above table is just the transpose of Table \ref{tab:chart}. Formally, we have
\begin{equation}
\ket{l}_\text{oam}\ket{k}_\text{path} \xrightarrow{G^{d_r}_{d_s}} \ket{l}_\text{oam}\ket{\lfloor\frac{k\oplus l}{d_s}\rfloor}_\text{path} \nonumber	
\end{equation}

\end{document}